\documentclass[aps,prd,twocolumn,groupedaddress,amssymb,eqsecnum,showpacs,epsfig]{revtex4}
\usepackage{graphicx}
\usepackage{bm}
\usepackage{dcolumn}
\usepackage{amsmath}
\def \lleq {\lower0.9ex\hbox{ $\buildrel < \over \sim$} ~}
\def \ggeq {\lower0.9ex\hbox{ $\buildrel > \over \sim$} ~}

\def\pa{\partial}
\def\bq{{\bf q}}
\def\bqd{{\bf \dot{q}}}
\def\qd{\dot{q}}
\def\alp{\alpha}

\newcommand{\rd}{{\rm d}}

\def \orad   {\Omega_{0r}}
\def \ol   {\Omega_{\Lambda}}
\def \ode   {\Omega_{DE}}

\def \omms   {\Omega_m}

\def \omr   {\Omega_r}
\def \omm  {\Omega_{0 {\rm m}}}

\def \beq  {\begin{equation}}
\def \eeq  {\end{equation}}
\def \ber  {\begin{eqnarray}}
\def \eer  {\end{eqnarray}}

\begin{document}
\newcommand{\newc}{\newcommand}

\newc{\be}{\begin{equation}}
\newc{\ee}{\end{equation}}
\newc{\ba}{\begin{eqnarray}}
\newc{\ea}{\end{eqnarray}}
\newc{\bea}{\begin{eqnarray*}}
\newc{\eea}{\end{eqnarray*}}
\newc{\D}{\partial}
\newc{\ie}{{\it i.e.} }
\newc{\eg}{{\it e.g.} }
\newc{\etc}{{\it etc.} }
\newc{\etal}{{\it et al.}}
\newcommand{\nn}{\nonumber}
\newc{\ra}{\rightarrow}
\newc{\lra}{\leftrightarrow}
\newc{\lsim}{\buildrel{<}\over{\sim}}
\newc{\gsim}{\buildrel{>}\over{\sim}}
\title{Reconstruction of the Scalar-Tensor Lagrangian from a $\Lambda$CDM Background and Noether Symmetry}
\author{S. Capozziello$^a$, S. Nesseris$^b$ and L. Perivolaropoulos$^b$}
\email{leandros@uoi.gr}

\affiliation{$^a$ Dipartimento di Scienze Fisiche, Universit\`a di
Napoli "Federico II", and INFN, Sez. di Napoli, Compl. Univ. di
Monte S. Angelo, Edificio G, Via Cinthia, I-80126 - Napoli, Italy
\\
$^b$ Department of Physics, University of Ioannina, Greece}
\date{\today}

\begin{abstract}
We consider scalar-tensor theories and reconstruct their potential
$U(\Phi)$ and  coupling $F(\Phi)$ by demanding a background
$\Lambda$CDM cosmology. In particular we impose a background cosmic
history $H(z)$ provided by the usual flat $\Lambda$CDM
parameterization through the radiation ($w_{eff}=1/3$), matter
($w_{eff}=0$) and deSitter ($w_{eff}=-1$) eras. The cosmological
dynamical system which is constrained to obey the $\Lambda$CDM
cosmic history  presents five critical points in each era, one of
which corresponding to the standard General Relativity (GR).  In the
cases  that differ from GR, the reconstructed coupling and potential
are of the form $F(\Phi)\sim \Phi^2$ and $U(\Phi)\sim F(\Phi)^m$
where $m$ is a constant. This class of scalar tensor theories is
also theoretically motivated by a completely independent approach:
imposing maximal Noether symmetry on the scalar-tensor Lagrangian.
This approach provides independently: $i)$ the form of the coupling
and the potential as $F(\Phi)\sim \Phi^2$ and $U(\Phi)\sim
F(\Phi)^m$, $ii)$ a conserved charge related to the potential and
the coupling and $iii)$ allows the derivation of exact solutions by
first integrals of motion.
\end{abstract}
\maketitle

\section{Introduction}
There is accumulating observational evidence based mainly on Type
Ia supernovae standard candles \cite{SN} and also on standard
rulers \cite{CMB,BAO} that the universe has entered a phase of
accelerating expansion at a recent cosmological timescale. Such an
expansion implies the existence of a repulsive pressure on
cosmological scales which counterbalances the gravitational
attraction  of matter on these scales giving rise to an overall
accelerating behavior. There have been several theoretical
approaches (see \cite{Copeland:2006wr,review} for a review) to
understand this phenomenon. The simplest of  such approaches
assumes the existence of a positive cosmological constant which is
small enough to have started dominating the universe at recent
times. The predicted cosmic expansion history, in this case,
(assuming flatness) is \be H(z)^2 = \left(\frac{{\dot
a}}{a}\right)^2 = H_0^2 \left[\omm (1+z)^3 + \orad (1+z)^4 +
\ol\right] \label{lcdmz} \ee where
$\orad=\frac{\rho_r}{\rho_{crit}}\simeq 10^{-4}$ is the energy
density of radiation today normalized over the critical density
for flatness $\rho_{crit}$. Also
$\omm=\frac{\rho_{m}}{\rho_{crit}}\simeq 0.3$ is the normalized
present matter density and $\ol=1-\omm-\orad$ is the normalized
energy density due to the cosmological constant. This model
provides an excellent fit to the cosmological observational data
\cite{CMB} and has the additional bonus of simplicity due to a
single free parameter. Despite its simplicity and good fit to the
data, such a model fails to explain why the cosmological constant,
 so unnaturally small dominates the universe at
recent cosmological times. This  problem is known as the {\it
coincidence problem}. Furthermore, there is no urgent theoretical
reason implying  $\ol\simeq 0.7$ and $\omm\simeq 0.3$ at present
time so also a {\it fine tuning problem} has to be taken into
account.

In the effort to address these problems, several models, which
essentially can be grouped in two classes,  have been proposed: The
first class assumes that General Relativity (GR) is  valid at
cosmological scales and attributes the accelerating expansion to a
{\it dark energy} component which has repulsive gravitational
properties due to its negative pressure. The role of dark energy is
usually played by a scalar field  minimally coupled to gravity
called {\it quintessence} \cite{quin}. Alternatively, the role of
dark energy can be played by various perfect fluids (eg Chaplygin
gas \cite{chapgas}) topological defects \cite{defectsde},
holographic dark energy \cite{holographic} etc.

The second class of models attributes the accelerating expansion
to  modifications and extensions of GR  which converts gravity to
a repulsive interaction at late times and on cosmological scales.
Examples of this class of models include scalar-tensor theories
\cite{stensor,Gannouji}, $f(R)$ extended gravity theories
\cite{fRpapers}, braneworld models \cite{braneworld} etc. An
advantage of models in this class is that they naturally allow
\cite{Perivolaropoulos:2005yv,BEPS00}  a super-accelerating
expansion of the universe where the effective dark energy equation
of state $w=\frac{p}{\rho}$ crosses the phantom divide line
$w=-1$. Such a crossing is consistent with current cosmological
data \cite{Alam:2003fg,Nesseris:2006er}.

A representative model of the second class is provided by
scalar-tensor theories of gravity. In these theories, the Newton
constant is obtained by  dynamical properties expressed through
the coupling $F(\Phi)$. Dynamics is given by the Lagrangian
density
\cite{Cap1993,Cap1994,Boisseau:2000pr,Esposito-Farese:2000ij} \be
{\cal L}=\frac{F(\Phi)}{2}~R - \frac{1}{2}~\epsilon~g^{\mu\nu}
\partial_{\mu}\Phi
\partial_{\nu}\Phi
- U(\Phi)  + {\cal L}_{\rm m}[\psi_{\rm m}; g_{\mu\nu}]\
\label{actfr} \ee where ${\cal L}_{\rm m}[\psi_{\rm m};
g_{\mu\nu}]$ represents matter fields approximated by a
pressureless perfect fluid in the dust dominated regime. We have
set $8\pi G=1$ and $\epsilon=\pm 1$ for standard scalar and 
phantom fields respectively, i.e. we have negative kinetic energy
for the scalar degree of freedom. Note, however that negative
energy needs $\epsilon = -1$, but there also exist positive energy
configurations with $\epsilon = -1$. This happens due to the fact
that the kinetic term of the actual spin-0 degree of freedom does
not come only from the obvious $(\partial_{\mu} \Phi)^2$, but also
from the cross term $F(\Phi) R$, since $R$ involves second
derivatives (see \cite{Esposito-Farese:2000ij} for details). As
discussed below, for $\epsilon=0$ and $\Phi\rightarrow R$ the
Lagrangian (\ref{actfr}) can also describe $f(R)$ generalizations
of GR.

In the present study, we reconstruct the potential $U(\Phi)$ and
the coupling $F(\Phi)$. Instead of specifying various forms of
$U(\Phi)$ and $F(\Phi)$ and finding the corresponding cosmological
dynamics, we specify the cosmological dynamics to that of the
$\Lambda$CDM cosmology and search for  possible corresponding
forms of $U(\Phi)$ and $F(\Phi)$. The original method for the
reconstruction of scalar-tensor theories from a given cosmic
history $H(z)$ was introduced in Ref. \cite{BEPS00} and applied to
specific cases of late cosmic history in Ref.
\cite{Esposito-Farese:2000ij,Perivolaropoulos:2005yv}. Our
reconstruction approach is different in two aspects:
\begin{itemize} \item We use a dynamical system formalism and find
the critical points that determine the generic evolution of the
system. \item We start the reconstruction from the radiation era
rather than focusing only at late times through the acceleration
epoch. \end{itemize} In particular, we consider the general
dynamical system for scalar-tensor theories and study the dynamics
of $U(\Phi)$ and $F(\Phi)$ using as input a $\Lambda$CDM cosmic
expansion history. Our study is performed both analytically (using
the critical points and their stability) and numerically by
explicitly solving the dynamical system.

Since our reconstruction assumes a fixed cosmic history background
the stability of the critical points we find should be interpreted
with care. The fixing of the cosmic history has eliminated
perturbations of $H(z)$ but has allowed for perturbations in the
forms of the coupling $F(\Phi)$ and the potential $U(\Phi)$. In the
usual stability approach the coupling $F(\Phi)$ and potential
$U(\Phi)$ are fixed and perturbations are allowed in the cosmic
history $H(z)$ to determine the stability of the phase space
trajectories. Since the later stability approach is more physical
(but does not lead to a reconstruction) we focus on the actual
values of the critical points and interpret their stability only as
a test of the corresponding numerical evolution of the same system.
Thus, all the critical points we find are assumed equally important
cosmologically, independent of their stability.

The structure of the paper is the following: In the next section we
derive the dynamical system for the cosmological dynamics of
scalar-tensor theories. Using as input a particular cosmic history
$H(z)$ (eg $\Lambda$CDM), we show how  this system can be
transformed so that its solution provides the dynamics and the
functional form of $U(\Phi)$ and $F(\Phi)$. We also study the
dynamics of this transformed system analytically by deriving its
critical points during the three eras of the cosmic background
history (radiation, matter and deSitter). We find that the
cosmological dynamical system constrained to obey the $\Lambda$CDM
cosmic history has five critical points in each era, one of which
corresponds to GR. In Sect. III, we use the solution of the above
system to reconstruct the cosmological evolution and functional form
of the coupling and potential, which are of the form $F(\Phi)\sim
\Phi^2$ and $U(\Phi)\sim F(\Phi)^m$ where $m$ is an arbitrary
constant. We show that such forms are also motivated by a completely
independent approach ie by imposing Noether symmetry on the
scalar-tensor Lagrangian \cite{cimento}. We also demonstrate the
agreement between the analytical and numerical results of our
reconstruction scheme. Finally, in Sect.IV, we conclude, summarize
and refer to future prospects of this work.

\section{Dynamics of Scalar-Tensor Cosmologies}

Let us consider the action (\ref{actfr}) describing the dynamics
of Scalar Tensor theories in the Jordan frame. In the context of
flat Friedman-Robertson-Walker (FRW) universes, the metric is
homogeneous and isotropic \ie \be \rd s^{2}=-\rd
t^{2}+a^{2}(t)\,\rd{\bf x}^{2}\label{frwmet}\ee Variation of the
action (\ref{actfr}) with respect to the metric leads to the
following dynamical equations which are the generalized Friedman
equations \cite{Marek1995,Esposito-Farese:2000ij,Boisseau:2000pr}

\ba 3F H^2 &=&
\rho_{\rm m} + \rho_r + \frac{1}{2} \epsilon {\dot \Phi}^2 -3H{\dot F} + U \label{stfe1} \\
-2F{\dot H} & = & \rho_{\rm m} + \frac{4}{3}\rho_r + \epsilon
{\dot \Phi}^2 + {\ddot F} - H{\dot F} \label{stfe2} \ea and
variation with respect to $\Phi$ gives the Klein-Gordon equation:
\be \epsilon (\ddot{\Phi}+3 H \dot{\Phi})=3F(\Phi)_{,\Phi}(\dot
H+2 H^2)-U(\Phi)_{,\Phi} \label{klgord}\ee where we have assumed
the presence of perfect fluids $\rho_{\rm m}$, $\rho_r$
representing the matter and radiation energy densities which are
conserved according to \ba
 &  & \dot{\rho}_{{\rm m}}+3H\rho_{{\rm m}}=0\,,\label{mcon}\\
 &  & \dot{\rho}_{{\rm r}}+4H\rho_{{\rm r}}=0\,.\label{radcon}\ea

The background equations in eqs. (\ref{stfe1}) and (\ref{stfe2}) can
be rewritten in a more convenient form, which is easier to confront
with observations (see for example Ref. \cite{Tsujikawa:2007gd}):

\be 3 F_0 H^2=\rho_{DE}+ \rho_{\rm m} + \rho_r, \label{de1}\ee

\be -2 F_0 \dot{H}= \rho_{DE}+  p_{DE}+\rho_{\rm m} +
\frac{4}{3}\rho_r \label{de2} \ee

where we have set

\be \rho_{DE} = \frac{1}{2} \epsilon {\dot \Phi}^2 -3H^2(F-F_0)
-3H{\dot F} + U \ee

\be p_{DE} = \frac{\epsilon}{2} {\dot \Phi}^2 + {\ddot F} +2
H{\dot F}- U-(2 \dot{H} +3 H^2)(F_0-F)\ee

\noindent and the subscript ``0'' denotes present day values. The
function $\rho_{DE}$ defined in this way can be shown to satisfy
the usual energy conservation equation:

\be \dot{\rho}_{DE}+3 H (\rho_{DE}+p_{DE})=0 \ee

where the DE equation of state is defined as

\be w_{DE}\equiv \frac{p_{DE}}{\rho_{DE}}=-1+\frac{\epsilon {\dot
\Phi}^2 + {\ddot F} - H{\dot F}-2 \dot{H}(F_0-F)}{\frac{1}{2}
\epsilon {\dot \Phi}^2 -3H^2(F-F_0) -3H{\dot F} + U } \ee

By using eqs. (\ref{de1}) and (\ref{de2}) we can express the
equation of state $w_{DE}$ as

\be w_{DE}= - \frac{3E(z) -(1+z) (d E(z)/dz)}{3E(z) - 3
\Omega_m^{(0)}(1+z)^3} \ee where we have set $E(z)\equiv
H^2(z)/H_0^2$. This relation is exactly the same as in standard
Einstein gravity \cite{Copeland:2006wr} and therefore we can
constrain $w_{DE}$ from Type Ia supernovae observations in the
same way.

In order to study the cosmological dynamics implied by equations
(\ref{stfe1}), (\ref{stfe2}) and (\ref{klgord}) we express them as
a dynamical system of first order differential equations. To
achieve this, let us first  write (\ref{stfe1}) in dimensionless
form as \be 1=\frac{\rho_{\rm m}}{3FH^2}+\frac{\rho_r}{3FH^2} +
\epsilon \frac{\Phi^{'2}}{6F}+\frac{U}{3FH^2}-\frac{F'}{F}
\label{acon} \ee where  \be '=\frac{d}{d{\rm
ln}a}\equiv\frac{d}{dN}=\frac{1}{H}\frac{d}{dt} \label{prdef} \ee
Let us now define the dimensionless variables $x_1,...,x_4$ as \ba
x_{1} & = & -\frac{F'}{F}\,,\label{x1}\\
x_{2} & = & \frac{U}{3FH^{2}}\,,\label{x2}\\
x_{3}^2 & = & \frac{\Phi^{'2}}{6F}\,,\label{x3}\\
x_{4} & = & ~\frac{\rho_{{\rm r}}}{3FH^{2}}=\omr \,.\label{x4}\ea
where we can associate $x_4$ with $\omr$ and $x_1+x_2+\epsilon
x_3^2 \equiv \ode$ with curvature dark energy (dark gravity).
Defining also $\omms\equiv \frac{\rho_{\rm m}}{3FH^2}$ we can
write equation (\ref{acon}) as \be \omms=1-x_1-x_2-\epsilon
x_3^2-x_4 \label{acon1} \ee Let us now use (\ref{prdef}) to
express (\ref{stfe2}) as \be \frac{H'}{H}=-\frac{\rho_{\rm
m}}{2FH^2}-\frac{2}{3}\frac{\rho_{r}}{FH^2}-\epsilon
\frac{\Phi^{'2}}{2F}-\frac{F''}{2F}-\frac{H'F'}{2 F
H}+\frac{F'}{2F} \label{bau1} \ee or \be x_1'=3-2x_1-3x_2+x_4+3
\epsilon x_3^2+x_1^2+2 \frac{H'}{H}-x_1 \frac{H'}{H}\label{au1}
\ee Differentiating $x_4$ of (\ref{x4}) with respect to $N$, we
have \be
x_4'=\frac{\rho_{r}'}{3FH^2}-\frac{\rho_{r}}{3FH^2}\frac{F'}{F}-\frac{2\rho_{r}}{3FH^2}\frac{H'}{H}
\label{bau2} \ee or \be x_4'=-4x_4+x_4 x_1-2 x_4 \frac{H'}{H}
\label{au4} \ee where we have  used (\ref{radcon}). Similarly,
differentiating (\ref{x2}) with respect to $N$, we find \be
x_2'=x_2\left[x_1(1-m)-2\frac{H'}{H}\right] \label{au2}\ee where
\be m\equiv \frac{U_{,\Phi}/U}{F_{,\Phi}/F} \label{mdef} \ee and
$,_\Phi$ implies derivative with respect to $\Phi$. Finally
differentiating (\ref{x3}) with respect to $N$ and using
(\ref{klgord}),  one finds \be \epsilon (x_3^2)'=\epsilon x_3^2
x_1-6 \epsilon x_3^2-2 x_1+m x_2 x_1 -2 \epsilon x_3^2
\frac{H'}{H}-x_1 \frac{H'}{H} \label{au3} \ee The dynamical system
(\ref{au1}), (\ref{au4}), (\ref{au2}), (\ref{au3}) describes the
cosmological dynamics of Scalar-Tensor theories. In two special
limits, it can be transformed into the dynamical systems obtained
in $f(R)$ theories and Quintessence respectively \cite{CapOd}.
However, we have to say that there is one more degree freedom in
scalar-tensor theories, compared to $f(R)$ theories, since here we
have two arbitrary functions, i.e. $F(\Phi)$ and $U(\Phi)$. In
general, we should add a further parameter $n$ which could be
related to $F_{,\Phi}$ and  $F_{,\Phi\Phi}$, beside the above $m$.
In fact, if we do not try to reconstruct the function $F(\Phi)$,
such a function can be fixed a priori and the corresponding
parameter $n$ would be, for example, $F_{,\Phi}/F$, or some similar
condition. In such a case $H(N)$ would not be fixed as in our
reconstruction approach but would have to be determined by the
autonomous system. This is the approach followed in
\cite{SanteLeach} where $F_{\Phi}/F$ is not present as a variable in
the autonomous system since $F(\Phi)$ is fixed a priori.  The
parameters in \cite{SanteLeach} are the power law of the fixed
potential (there called $n$) and $\xi$, the coupling
$F(\Phi)=\xi\Phi^2$. In that case, $H(N)$ and its perturbations are
allowed to vary. However, in a reconstruction approach where $H(N)$
is fixed a priori, while $F_{,\Phi}/F$ (and its perturbations) are
allowed to vary,  $F(\Phi)$ can be reconstructed from the autonomous
system. Thus, in our reconstruction approach where $H(N)$ is fixed,
$n$ is no more a parameter and it is allowed to evolve. In this
sense, the present approach is more general than that presented in
\cite{SanteLeach}. Finally, as we shall see below, the reconstructed
function $F(\Phi)$ and $U(\Phi)$ can be independently obtained
asking for a sort of "first principle", the existence of a Noether
Symmetry. Since the results coincide (the form of $F$ and $U$ are
the same in both the approaches), we are confident that the
presented method is consistent.

Considering again the $f(R)$-theories, by setting
$U=\frac{FR-f}{2}$ \cite{Amendola:2006we} and using the
transformations

\begin{eqnarray}
  x_1 & \rightarrow & \tilde{x}_1 \\
  x_2 & \rightarrow & \tilde{x}_2+ \tilde{x}_3 \\
  x_4 & \rightarrow & \tilde{x}_4 \\
  \frac{H'}{H} & \rightarrow & \tilde{x}_3-2 \\
  \Phi & \rightarrow & R \\
  \epsilon & \rightarrow & 0
\end{eqnarray}
one recovers the dynamical system of $f(R)$ theories (eqs (2.15),
(2.17) and (2.21) in Ref. \cite{Fay:2007uy} ), where the
\textit{tilded} ($\tilde{\;\;}$) quantities are the ones defined
for $f(R)$ theories:

\be
\tilde{x}_1'=-1-\tilde{x}_3-3\tilde{x}_2+\tilde{x}_1^2+\tilde{x}_4
\label{frau1} \ee \be
\tilde{x}_2'=-\tilde{x}_3'-2\tilde{x}_3(\tilde{x}_3-2)-\tilde{x}_2(2\tilde{x}_3-\tilde{x}_1-4)
\label{frau2a} \ee \be
\tilde{x}_4'=-2\tilde{x}_3~\tilde{x}_4+\tilde{x}_1~\tilde{x}_4
\label{frau4} \ee

On the other hand, the following set of transformations gives the
autonomous system for Quintessence (see eqs (175) and (176) in
Ref. \cite{Copeland:2006wr}):

\begin{eqnarray}
  x_1 & \rightarrow & 0 \\
  x_2 & \rightarrow & y^2 \\
  x_3^2 & \rightarrow & x^2 \\
  x_4 & \rightarrow & 0
\end{eqnarray}
with dynamical equations  \be x'=-3x+\frac{\sqrt{6}}{2} \epsilon
\lambda y^2+\frac{3}{2} x ( \epsilon x^2 +1-y^2) \label{quinteqs1}
\ee

\be y'=-\frac{\sqrt{6}}{2} \lambda x y +\frac{3}{2} y ( \epsilon
x^2 +1-y^2) \label{quinteqs2} \ee and $\lambda = - U_{,\Phi}/U$.

\vspace{0pt}
\begin{table*}[t!]
\begin{center}
\caption{The critical points of the system (\ref{au1}),
(\ref{au4}), (\ref{au2}), (\ref{au3}) and their eigenvalues in
each one of the three eras (rad. era $N<-{\rm
ln}\frac{\omm}{\orad} $, matter era $-{\rm
ln}\frac{\omm}{\orad}<N<-\frac{1}{3}{\rm ln}\frac{\ol}{\omm} $,
deSitter era $N>-\frac{1}{3}{\rm ln}\frac{\ol}{\omm} $).
\label{table1}}
\begin{tabular}{ccccccccc}
\hline \hline\\
\vspace{6pt} \textbf{Era} &  \textbf{CP}  &\hspace{7pt} \textbf{$x_1$} \hspace{7pt}& \hspace{7pt} \textbf{$x_2$} \hspace{7pt}& \hspace{7pt} \textbf{$x_3^2$}  \hspace{7pt}& \hspace{7pt} \textbf{$x_4$} \hspace{7pt}& \hspace{7pt} \textbf{$\Omega_m$} \hspace{7pt} & \textbf{$\Omega_{DE}$} \hspace{7pt}& \hspace{7pt} \textbf{Eigenvalues} \hspace{7pt} \\
\hline
\vspace{6pt}                      &   $R_1$ &   2     &   0    &      -1    &    0   &   0     &     1    &(2,3,1,6-2$m$)    \\
\vspace{6pt}                      &   $R_2$ &   1     &   0    &       0    &    0   &   0     &     1    &(1,2,-1,5-$m$)    \\
\vspace{6pt} \textbf{Radiation}   &   $R_3$ &  -1     &   0    &       0    &    0   &   2     &    -1    &(-1,-2,-3,3+$m$)  \\
\vspace{6pt} $w_{eff}=\frac{1}{3}$&   $R_4$ & $\frac{4}{-1+m}$ & $\frac{15-8m+m^2}{3(m-1)^2}$ & $\frac{2(m-5)m}{3(m-1)^2}$ & 0 & 0     &     1 & $(\frac{4}{m-1},\frac{m+3}{m-1}$, see \footnote[1]{\; the other two eigenvalues are:$\;\; -\frac{3 m+\sqrt{8 m^3-63 m^2+118 m+1}-11}{2 (m-1)},\frac{-3 m+\sqrt{8 m^3-63 m^2+118 m+1}+11}{2 m-2} $})\\
\vspace{6pt}                      &   $R_5$ & 0       &   0    &       0    &    1   &   0     &     0    & (1,-1,-2,4)\\
\hline
\vspace{6pt} & $M_1$&  2  &  0   &  -1     & 0   & 0     &     1 & (1,2,1/2,$5-2m$) \\
\vspace{6pt} & $M_2$&  3/2  &  0  &  -1/2  & 0   & 0     &     1 & (1/2,3/2,-1/2,$-3/2(m-3)$) \\
\vspace{6pt} \textbf{Matter}     & $M_3$ & 0  & 0 & 0 & 0  & 0     &     0 & (-1,-3/2,-2,3)\\
\vspace{6pt} $w_{eff}=0$& $M_4$ & $\frac{3}{m-1}$ & $\frac{15-11m+2m^2}{4(m-1)^2}$ & $\frac{1-9m+2m^2}{4(m-1)^2}$ & 0 & 0     &     1 & $(\frac{4-m}{m-1},\frac{3}{m-1}$, see \footnote[2]{\; the other two eigenvalues are:$\;\; -\frac{7 m+\sqrt{48 m^3-263 m^2+358 m+1}-19}{4 (m-1)},\frac{-7 m+\sqrt{48 m^3-263 m^2+358 m+1}+19}{4 m-4}$})\\
\vspace{6pt}  & $M_5$ & 1 & 0 & -1/4 & 1/4 & 0     &     3/4 & (1,-1/2,-1,$4-m$)\\
\hline
\vspace{6pt}  & $\Lambda_1$ &   2   &  0   &  -1   &   0 &  0     &     1   & (-2,-1,-1,$2-2m$)\\
\vspace{6pt} & $\Lambda_2$ &   3   &  0   &  -2   &   0  &  0     &     1   & (-1,0,1,$3-3m$)  \\
\vspace{6pt} \textbf{deSitter}  & $\Lambda_3$\footnote[3]{\; Notice that $\Lambda_2$ and $\Lambda_3$ are degenerate} & 3 & 0 & -2 & 0 & 0     &     1 & (-1,0,1,$3-3m$) \\
\vspace{6pt} $w_{eff}=-1$& $\Lambda_4$ & 0  & 1 & 0 & 0  & 0     &     1 & $\left(-4,-3,-\frac{\sqrt{24 m+1}+5}{2} ,\frac{\sqrt{24 m+1}-5}{2} \right)$\\
\vspace{6pt}  & $\Lambda_5$ & 4  & 0 & -4 & 1 & 0     &     0& (1,1,2,$4-4m$) \\
\hline \hline
\end{tabular}
\end{center}
\end{table*}

It is worth noting that the results of our analysis do not rely on
the use of any particular form of $H(z)$. They only require that
the universe goes through the radiation era (high redshifts),
matter era (intermediate redshifts) and acceleration era (low
redshifts). The corresponding total effective equation of state
\be w_{eff}=-1-\frac{2}{3}\frac{H'(N)}{H(N)} \label{weff}
\ee is \ba w_{eff}&=&\frac{1}{3} \;\;\; {\rm Radiation\; Era} \nn \\
 w_{eff}&=&0 \;\;\;\; {\rm Matter\; Era} \label{eraweff} \\
w_{eff}&=&-1 \;\;\; {\rm deSitter\; Era} \nn \ea For the sake of
definiteness however, we will assume a specific form for $H(z)$
corresponding to a $\Lambda$CDM cosmology (\ref{lcdmz}) which, in
terms of $N$, takes the form \be H(N)^2=H_0^2 \left[ \omm e^{-3N}+
\orad e^{-4N}+\ol\right] \label{lcdmn} \ee where $N\equiv {\rm ln}
a=-{\rm ln}(1+z)$ and $\ol=1-\omm-\orad$. Also, we can use
(\ref{lcdmn}) to find $\frac{H'(N)}{H(N)}$, a quantity needed in
the dynamical system, as \be \frac{H'(N)}{H(N)}=\frac{-3\,{{\Omega
}_{0m}}e^{-3 N} - 4\,{{\Omega }_{0r}}e^{-4 N}}{2\,\left( 1 -
{{\Omega }_{0m}} - \ {{\Omega }_{0r}}+ {{\Omega }_{0m}}e^{-3 N}  +
{{\Omega }_{0r}}e^{-4 N} \right) }\label{hpoverh} \ee The crucial
generic properties of $\frac{H'(N)}{H(N)}$ are their values at the
radiation, matter and deSitter eras:

\ba \frac{H'(N)}{H(N)}&=& -2 \;\; \;\; N<N_{rm}
\label{hpoverhrad} \\
\frac{H'(N)}{H(N)}&=& -\frac{3}{2} \;\; \;\; N_{rm}<N<N_{m\Lambda}
\label{hpoverhmat}
\\\frac{H'(N)}{H(N)}&=& 0 \;\; \;\; N>N_{m\Lambda} \label{hpoverhds} \ea where $N_{rm}\simeq -{\rm
ln}\frac{\omm}{\orad}$ and $N_{m\Lambda}\simeq -\frac{1}{3}{\rm
ln}\frac{\ol}{\omm}$ are the $N$ values for the radiation-matter
and matter-deSitter transitions. For $\omm=0.3$, $\orad=10^{-4}$
we have $N_{rm}\simeq -8$, $N_{m\Lambda}\simeq -0.3$. The
transition between these eras is model dependent but rapid and it
will not play an important role in our analysis.

It is straightforward to study the dynamics of the system
(\ref{au1}), (\ref{au4}), (\ref{au2}), (\ref{au3}) by finding the
critical points and their stability in each one of the three eras.
Notice that even though this dynamical system is not autonomous at
all times, it can be approximated as such during the radiation,
matter and deSitter eras when $\frac{H'(N)}{H(N)}$ is
approximately constant. The critical points and their eigenvalues
are shown in Table I. An interesting feature to observe in Table I
is that in each era there are five critical points but only one of
them is a stable ``attractor" for a given value of $m$. Also, 
remembering that the positivity of the energy of the (helicity
zero) scalar partner of the graviton, i.e. the positivity of the
kinetic energy of the scalar field (see Ref. \cite{Gannouji}) is
expressed by:

\be \phi'^2=\frac{3}{4} \left(\frac{F'}{F}\right)^2+\frac{\epsilon
\Phi'^2}{2F}>0 \ee

then we have in dimensionless variables

\be \frac{x_1^2}{4}+x_3^2>0 \label{constr} \ee

All points of Table I satisfy (\ref{constr}) except $R_4$ and
$M_4$. In order for $R_4$ to satisfy (\ref{constr}), we must have
$m<2$ or $m>3$ while for $M_4$ it is $m<2$ or $m>5/2$. Therefore
the allowed region for $m$, so that we have critical points that
are physical ($\frac{x_1^2}{4}+x_3^2>0$), is \be m \leq 2 ~~~ or
~~~ m \geq 3 \label{rangem} \ee

Now, regarding the ``attractor" behavior of the system in each era
separately, we see that in:
\begin{itemize}
    \item Radiation Era: $R_3$ is an ``attractor" for $m<-3$ while $R_4$ is
    an ``attractor" for $-3<m<1$;
    \item Matter Era: $M_4$ is an ``attractor" for $m<1$;
    \item deSitter Era: $\Lambda_1$ is an ``attractor" for $m>1$ and $\Lambda_4$ is an ``attractor" for
    $m<1$.
\end{itemize}

As discussed in the introduction, the significance of the
``attracting" nature of the critical points is limited because the
reconstruction method we are using has forced us to eliminate
physical perturbations of $H(N)$ and has introduced unphysical
perturbations of the functional forms of the potential $U(\Phi)$ and
coupling $F(\Phi)$. However, we can use the ``attractor" critical
points as a prediction for the numerical evolution of our
reconstruction dynamical system.

A particularly interesting feature of the critical points of Table
I is that in all the cases that differ from GR, the expansion rate
in each era is induced by dark energy ($\Omega_{DE}=1$) which
implies that the scalar field $\Phi$ could also play the role of
dark matter if it is found to have the proper perturbation
properties at early times. We postpone the analysis of such
perturbations for a future study.

To confirm the dynamical evolution implied by the ``attractors" of
Table I, we have performed a numerical analysis of the dynamical
system (\ref{au1}), (\ref{au4}), (\ref{au2}) and (\ref{au3}) using
the ansatz (\ref{hpoverh}) for $\frac{H'(N)}{H(N)}$ with $\omm=0.3$
and $\orad=10^{-4}$. This ansatz, for $\frac{H'(N)}{H(N)}$, leads to
the $w_{eff}(N)$ shown in Fig. 1. We have set up the system
initially, on $R_4$, with $m=-0.5$. As seen in Fig. 2 and Fig. 3 the
system follows the evolution of $R_4$ from the Radiation Era, to the
Matter Era ($M_4$) and finally to the deSitter Era ($\Lambda_4$). We
have checked that if we choose initial conditions not exactly
coinciding with any of the other critical points then the system is
captured by the $R_4$ ``attractor" and follows the trajectory
mentioned above, i.e. $ (init) \rightarrow R_4 \rightarrow M_4
\rightarrow \Lambda_4$.

Finally, if the initial conditions of the system are set to be
exactly on the top of  other critical points then the system will
end up on the $\Lambda_1$ critical point.

The choice of a constant $m$ is justified by the fact that $x_1$
and $x_3^2$ are constants in each era (see Fig. 2). Also, the
potential $U(\Phi)$ and coupling $F(\Phi)$ that are mostly used in
the literature are power-laws or exponentials, which also give a
constant $m$ (see equation (\ref{mdef})). This feature will be
obtained also by Noether symmetries, as we will discuss below.

\begin{figure}[!t]
\hspace{0pt}\rotatebox{0}{\resizebox{.5\textwidth}{!}{\includegraphics{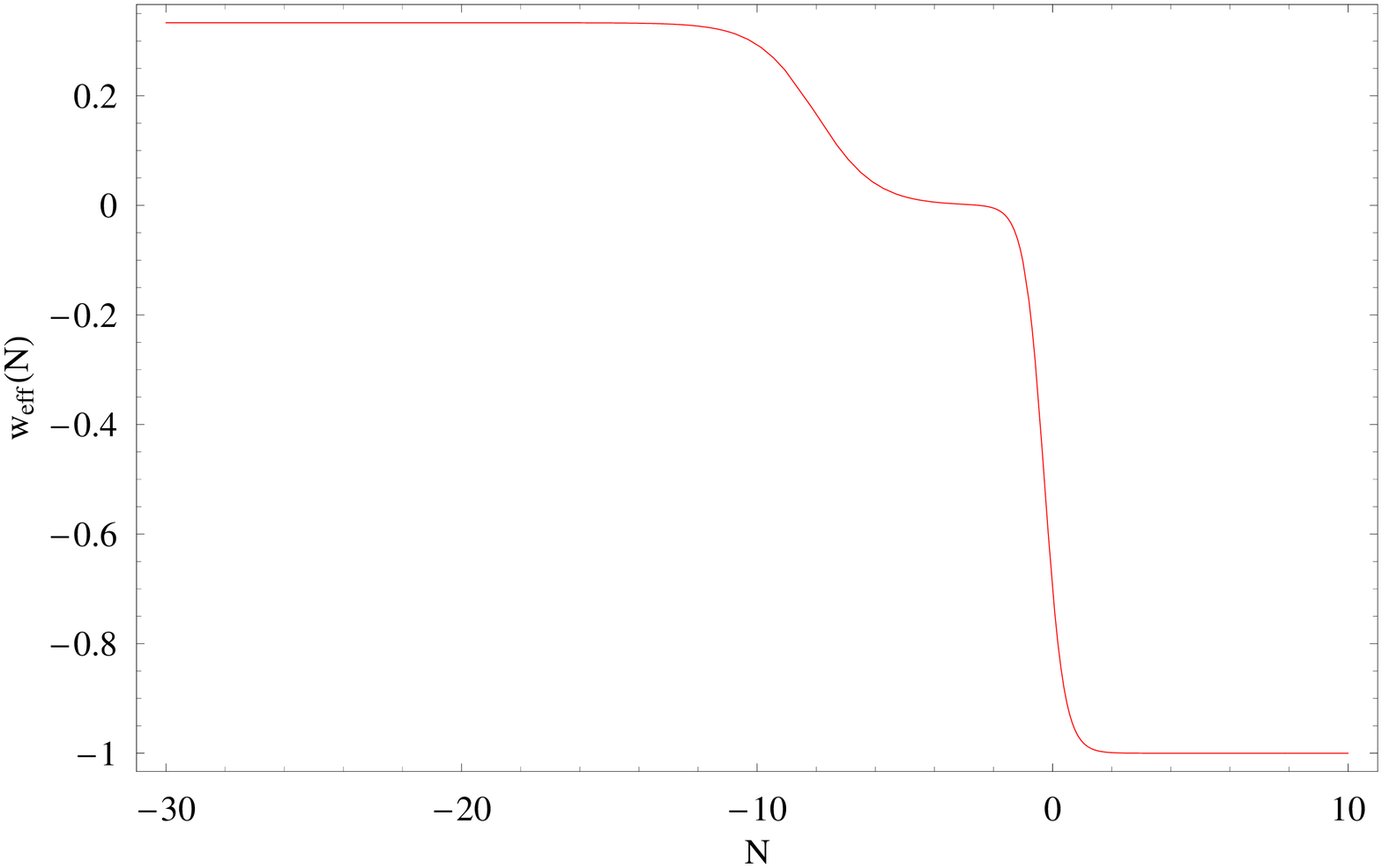}}}
\vspace{-20pt}{\caption{The effective equation of state
$w_{eff}(N)$ imposed on the dynamical system (obtained from
(\ref{weff}) using (\ref{lcdmn})).}} \label{fig1}
\end{figure}

\section{Reconstruction of $F(\Phi)$ and $U(\Phi)$}

\subsection{Analytical Results}

Our goal is now the reconstruction of the form of the potential
$U(\Phi)$ and coupling $F(\Phi)$  corresponding to each one of the
critical points of the system shown in Table I. Let us consider a
critical point of the form $({\bar x}_1,{\bar x}_2,{\bar
x}_3^2,{\bar x}_4)$. Using (\ref{x1}), we find \be F=F_0 e^{-{\bar
x}_1 N} \label{fsol1} \ee where $F_0=F(N=0)$ is the present value of
$F$. Using now eq. (\ref{x3}) we find

\ba \Phi(N)&=&-2\sqrt{6}~ \frac{{\bar x}_3}{{\bar x}_1} ~
F_0^{1/2} e^{-{\bar x}_1 N/2} +C \nn \\ &=& 2\sqrt{6}~ \frac{{\bar
x}_3}{{\bar x}_1}~ F_0^{1/2} \left(1-e^{-{\bar x}_1
N/2}\right)+\Phi_0 \label{PhiN} \ea

where \be C=\Phi_0 +2\sqrt{6}\; \frac{{\bar x}_3}{{\bar x}_1} \;
F_0^{1/2} \label{constC} \ee

and $\Phi_0\equiv \Phi(N=0)$. Equations (\ref{fsol1}) and
(\ref{PhiN}) allow us to eliminate N in favour of $\Phi$

\be F(\Phi)=\frac{1}{24}\frac{{\bar x}_1^2}{{\bar
x}_3^2}(\Phi-C)^2\equiv \xi  (\Phi-C)^2\label{FofPhi} \ee where
$\xi \equiv \frac{1}{24}\frac{{\bar x}_1^2}{{\bar x}_3^2}$. The
quadratic form of $F(\Phi)$ can be achieved in a completely
different approach imposing Noether symmetry in the scalar-tensor
Lagrangian.

From eq. (\ref{x2}), we have \be U(N)={\bar x}_2 \cdot 3F(N)
H(N)^2 \label{VofN} \ee

Using now the input form of $H(N)$ (Eq.(\ref{lcdmn})), we find the
dominant term of $H(N)$ in each era, that is

$ H(N)^2/H_0^2=\left\{%
\begin{array}{ll}
    \orad e^{-4N}, & \hbox{Rad. Era} \\
    \omm e^{-3N}, & \hbox{Mat. Era} \\
    1-\orad-\omm, & \hbox{dS Era} \\
\end{array}%
\right.$

Thus using eq.(\ref{PhiN}) we have $H(\Phi)$:

$H(\Phi)^2/H_0^2=\left\{%
\begin{array}{ll}
    \frac{\orad}{F_0^{4/{\bar x}_1}}\left[ \xi (\Phi-C)^2 \right]^{4/{\bar x}_1}, & \hbox{Rad. Era} \\
    \frac{\omm}{F_0^{3/{\bar x}_1}}\left[ \xi (\Phi-C)^2 \right]^{3/{\bar x}_1}, & \hbox{Mat. Era} \\
    1-\orad-\Omega_{\rm m}, & \hbox{dS Era} \\
\end{array}%
\right.$ and then we can  express (\ref{VofN}) in terms of $\Phi$
to get the relevant form \be U(\Phi)= \lambda
(\Phi-C)^{2+\alpha}\label{VofPhi} \ee

where,

$\lambda=\left\{%
\begin{array}{ll}
    3 {\bar x}_2 \frac{\orad}{F_0^{4/{\bar x}_1}} \xi^{1+4/{\bar x}_1}, & \hbox{Rad. Era} \\
    3 {\bar x}_2 \frac{\omm}{F_0^{3/{\bar x}_1}} \xi^{1+3/{\bar x}_1}, & \hbox{Mat. Era} \\
    3 {\bar x}_2 \xi (1-\orad-\omm), & \hbox{dS Era} \\
\end{array}%
\right.$

and,
$\alpha=\left\{%
\begin{array}{ll}
    8/{\bar x}_1, & \hbox{Rad. Era} \\
    6/{\bar x}_1, & \hbox{Mat. Era} \\
    0, & \hbox{dS Era} \\
\end{array}%
\right.$

By using eqs. (\ref{mdef}), (\ref{FofPhi}) and (\ref{VofPhi}), it
is straightforward to find  \be 2+\alpha = 2m \label{2am}\ee so
that Eq.(\ref{VofPhi}) can be written as \be U(\Phi)= \lambda
(\Phi-C)^{2m}\label{VofPhi1} \ee Notice that even though the
eigenvalue ${\bar x_1}$ changes in the sequence $R_4\rightarrow
M_4$, the exponent $2+\alpha$ remains constant.

\begin{figure*}[pt!]
\rotatebox{0}{\resizebox{1\textwidth}{!}{\includegraphics{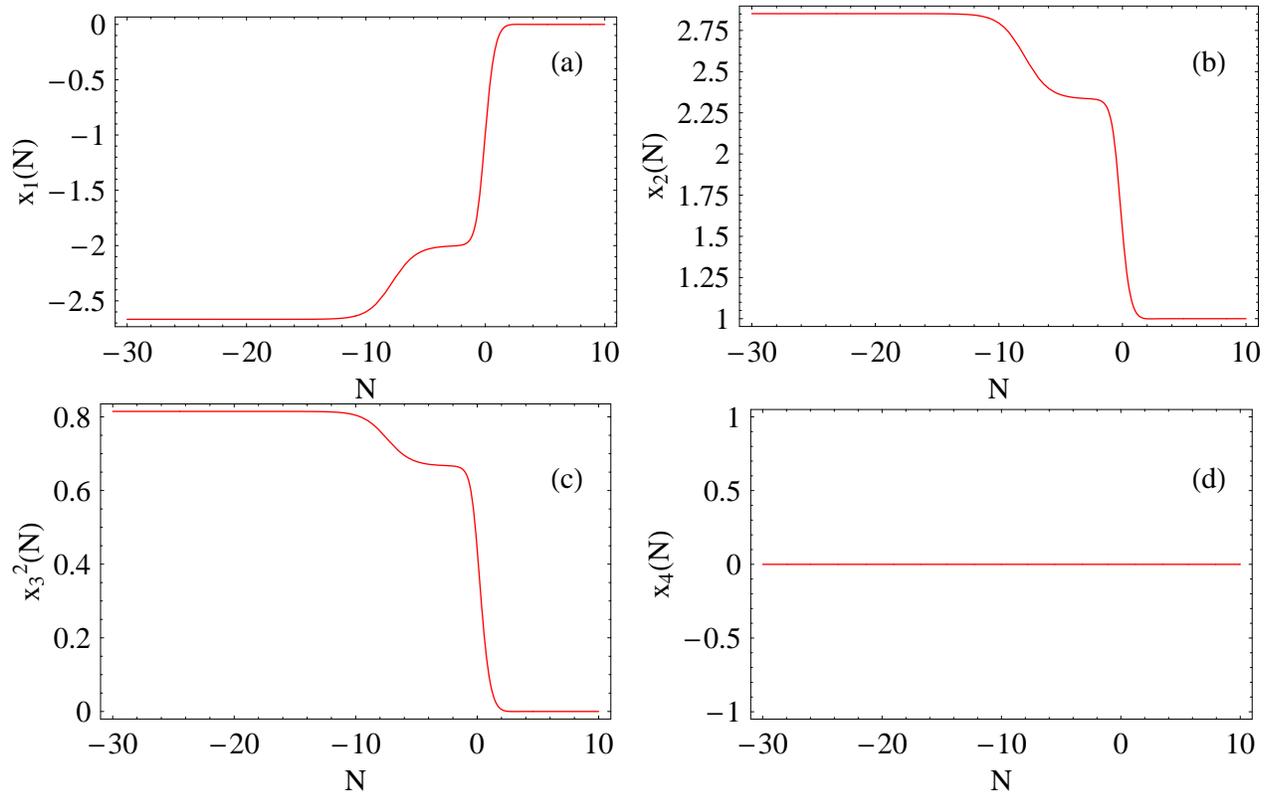}}}
\vspace{0pt}{ \caption{The evolution of the variables $x_1(N)$,
$x_2(N)$, $x_3^2(N)$ and $x_4(N)$. The system follows the
evolution of the ``attractor" through the three eras.}}
\label{fig2}
\end{figure*}

\begin{figure*}[pb!]
\rotatebox{0}{\resizebox{1\textwidth}{!}{\includegraphics{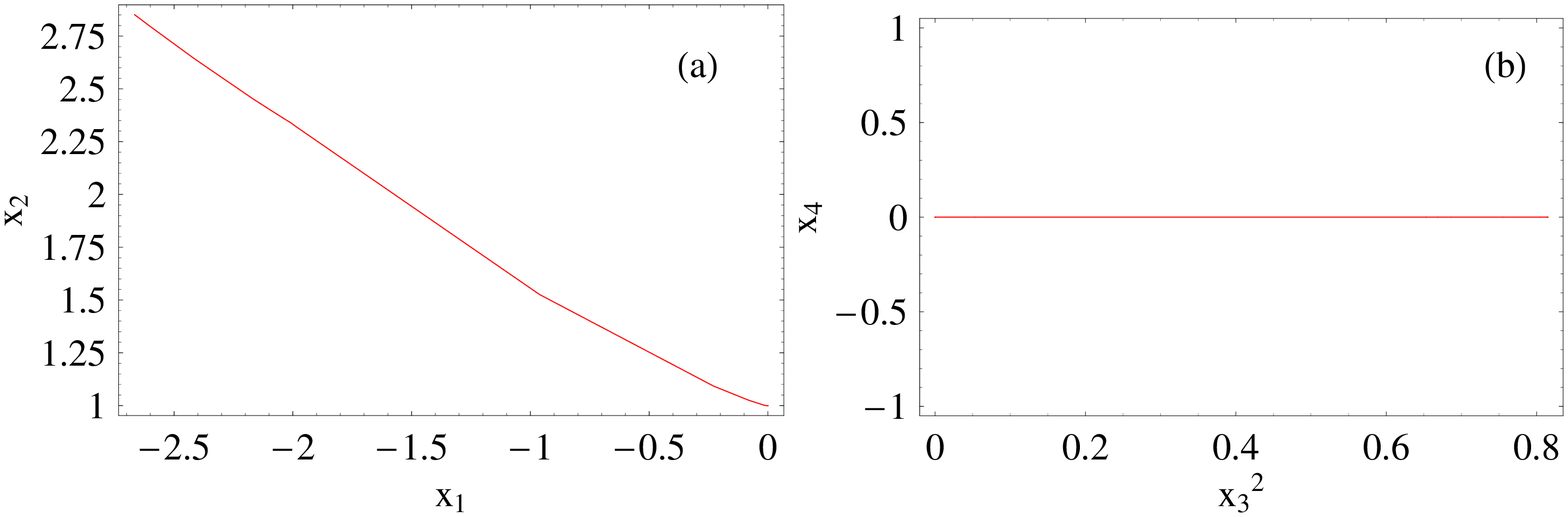}}}
\vspace{0pt}{ \caption{The phase space trajectories on  the $x_1 -
x_2$  plane (Fig3a) and $x_3^2 - x_4$ plane (Fig3b).}}
\label{fig3}
\end{figure*}
\clearpage

Furthermore, since $m$ is constant (see Eq.(\ref{2am}) )
Eq.(\ref{mdef}) allows to write $U$ in terms of $F$, i.e. \be U=c
F^m \label{UofF}\ee

The above analysis is only valid when the parameters ${\bar x}_1$,
${\bar x}_2$ and ${\bar x}_3^2$ are not equal to zero, as for
example is the case for $R_4$ or when we have perturbed the
initial conditions around a critical point. Otherwise, we have the
following cases:

\begin{itemize}
    \item ${\bar x}_2={\bar x}_3^2=0$, then $\Phi=\Phi_0=$\textit{const}, $F=F(N)$ given from eq. (\ref{fsol1}) and
    $U=0$, as is the case for $R_2$ and $R_3$.
    \item ${\bar x}_1={\bar x}_3^2=0$ and we are in the deSitter era,
    then $\Phi=\Phi_0=$\textit{const}, $F=F_0$ and
    $U=U_0$, as is the case for $\Lambda_4$.
    \item ${\bar x}_1={\bar x}_2={\bar x}_3^2=0$, then $\Phi=\Phi_0=$\textit{const}, $F=F_0$ and
    $U=0$, as is the case for $R_5$.

\end{itemize}

Note that if we were considering other (non-critical) points, not
only the reconstruction would involve a time dependence, but we
would not be able to reconstruct simultaneously $F(\Phi)$ and
$U(\Phi)$ from the single function (\ref{lcdmz}). Two arbitrary
functions obviously need two observed functions to be
reconstructed in the general case (see
\cite{Perivolaropoulos:2005yv,Boisseau:2000pr,Esposito-Farese:2000ij}).
In our case, however, the fact that the reconstruction occurs on
the critical points means that $m$ is fixed to a constant and this
closes the system of equations (\ref{au1}), (\ref{au4}),
(\ref{au2}), (\ref{au3}) allowing us to proceed with the
reconstruction numerically and analytically.

Also, it is easy to see that the reconstructed theories are merely
Brans-Dicke theories with an additional potential $U$. If we
define $F = \beta \phi_{BD} = \xi \Phi^2$ then $\Phi =
\sqrt{\frac{\beta}{\xi}\phi_{BD}}$ and the Lagrangian
(\ref{actfr}) becomes \be {\cal L}=\frac{\beta \phi_{BD}}{2}~R -
\frac{1}{2}~\frac{\omega_{BD}}{\phi_{BD}}~g^{\mu\nu}
\partial_{\mu}\phi_{BD}
\partial_{\nu}\phi_{BD}
- U(\phi_{BD})  + {\cal L}_{\rm m} \label{actbd} \ee where
$\omega_{BD} \equiv \epsilon \frac{\beta}{\xi}=constant $.

 A noteworthy feature of the above reconstruction scheme is
that Eq.(\ref{UofF}) along with Eq. (\ref{FofPhi}) are exactly the
conditions for the existence of a Noether symmetry in
Scalar-Tensor theories as we discuss in what follows.

\subsection{Noether Symmetries in Scalar-Tensor Gravity}

\subsubsection{Generalities on the method}

Solutions for the Lagrangian (\ref{actfr}) can be searched by the
so called Noether Symmetry Approach \cite{cimento}. This approach
allows, in principle, to find out cyclic variables related to
conserved quantities and then to reduce dynamics. Besides, the
existence of symmetries fixes the forms of the coupling $F(\Phi)$,
of the potential $U(\Phi)$ and gives the relation between them.

Let us give a quick summary of the approach for finite dimensional
dynamical systems before the application to our specific problem.

Let ${\cal L}(q^{i}, \dot{q}^i)$ be a canonical, non-degenerate
point-like Lagrangian in the configuration coordinates $q^i$ (the
``positions"), where \beq \label{01} \frac{\pa {\cal L}}{\pa
\lambda}=0\,;\;\;\;\;\;\;\; \mbox{det}H_{ij}\equiv \mbox{det}
\left|\left| \frac{\pa^2 {\cal L}}{\pa
\dot{q}^i\pa\dot{q}^j}\right|\right|\neq 0 \,. \eeq  $H_{ij}$ is the
Hessian matrix related to ${\cal L}$. The dot indicates derivatives
with respect to the affine parameter $\lambda$ which, in general,
corresponds to the time $t$. We are going to consider only
transformations which are point-transformations. Any invertible and
smooth transformation of the ``positions" $Q^{i}=Q^{i}({\bq})$
induces a transformation on the ``velocities" such that \beq
\label{04} \dot{Q}^i({\bq})=\frac{\pa Q^i}{\pa q^j}\qd^j\,. \eeq The
matrix ${\cal J}=|| \pa Q^i/\pa q^j ||$ is the Jacobian of the
transformation on the positions, and it is assumed to be nonzero.
The Jacobian $\widetilde{{\cal J}}$ of the ``induced" transformation
is easily derived and it has to be ${\cal J}\neq 0\rightarrow
\widetilde{{\cal J}}\neq 0$. Usually, this condition is not
satisfied in the whole space but only in the neighbor  of a point.
It is a {\it local transformation}. A point transformation
$Q^{i}=Q^{i}(\bq)$ can depend on one (or more than one) parameter.
In general, an infinitesimal point transformation is represented by
a generic vector field acting on the space $\{q^{i}, \dot{q}^i\}$.
The transformation induced by (\ref{04}) is then represented by \beq
\label{05} {\bf X}=\alp^{i}({\bq})\frac{\pa}{\pa q^{i}}+
\left(\frac{d}{d\lambda}\alp^{i}({\bq})\right)\frac{\pa}{\pa
\qd^i}\;. \eeq ${\bf X}$ is called the ``complete lift" of ${\bf X}$
\cite{morandi}. A function $f(\bq, \bqd)$ is invariant under the
transformation  ${\bf X}$ if \beq \label{06} L_{{\bf
X}}f\equiv\alp^{i}({\bq})\frac{\pa f}{\pa q^{i}}+
\left(\frac{d}{d\lambda}\alp^{i}({\bq})\right)\frac{\pa f}{\pa
\qd^i}\,=\,0\;, \eeq where $L_{{{\bf X}}}f$ is the Lie derivative of
$f$. In particular, if
\begin{equation} L_{{{\bf X}}}{\cal L}=0\,,\label{noether}\end{equation}
 ${\bf X}$ is said to be a {\it
symmetry} for the dynamics derived from the Lagrangian ${\cal L}$.
To see how Noether's theorem and cyclic variables are related, let
us consider a Lagrangian ${\cal L}$ and the related Euler-Lagrange
equations \beq \label{07} \frac{d}{d\lambda}\frac{\pa {\cal
L}}{\pa\qd^{j}}-\frac{\pa {\cal L}}{\pa q^{j}}=0\,. \eeq Let us
consider also the vector field (\ref{05}). By contracting (\ref{07})
with the $\alpha^{i}$'s, one obtains \beq \label{06a} \alp^{j}\left(
\frac{d}{d\lambda}\frac{\pa {\cal L}}{\pa \qd^j}- \frac{\pa {\cal
L}}{\pa q^j}\right)=0\,. \eeq Being \beq \label{06b}
\alp^{j}\frac{d}{d\lambda}\frac{\pa {\cal L}}{\pa \qd^j}=
\frac{d}{d\lambda}\left(\alp^j\frac{\pa {\cal L}}{\pa \qd
^j}\right)- \left(\frac{d \alp^j}{d\lambda}\right)\frac{\pa {\cal
L}}{\pa \qd ^j}\,, \eeq from (\ref{06a}), we have \beq \label{08}
\frac{d}{d\lambda}\left(\alp^{i}\frac{\pa {\cal L}}{\pa \qd^i}
\right)=L_{\bf X}{\cal L}\,. \eeq As a consequence,  the {\it
Noether Theorem} enunciates:

 If $L_{\bf X}{\cal L}=0$,  the function \beq \label{09}
\Sigma_{0}=\alp^{i}\frac{\pa {\cal L}}{\pa \qd^i} \,, \eeq is a
constant of motion.

It is worth noting that Eq.(\ref{09}) can be expressed independently
of coordinates as a contraction of ${\bf X}$ by a Cartan one-form
\beq \label{09a} \theta_{\cal L} \equiv \frac{\pa {\cal L}}{\pa
\qd^i}dq^i \; . \eeq  Thus Eq.(\ref{09}) can be written as \beq
\label{09b} i_{\bf X} \theta _{\cal L} = \Sigma_{0} \; . \eeq where
$i_{\bf X}$ is defined through the relation \beq i_{\bf X}dq^i =
\alpha ^i \eeq By a point--transformation, the vector field ${\bf
X}$ becomes \beq \label{09c} \widetilde{{\bf X}} = (i_{\bf X} d Q^k)
\frac{\pa}{\pa Q^k} +
     \left( \frac{d}{d\lambda} (i_{\bf X} d Q^k)\right) \frac{\pa}{\pa \dot{Q}^k} \; .
\eeq $\widetilde{{\bf X}}$ is still the lift of a vector field
defined on the ``space of positions". If ${\bf X}$ is a symmetry and
we choose a point transformation  such  that \beq \label{010} i_{\bf
X} dQ^1 = 1 \; ; \;\;\; i_{\bf X} dQ^i = 0 \;\;\; i \neq 1 \; , \eeq
we get \beq \label{010a} \widetilde{{\bf X}} = \frac{\pa}{\pa Q^1}
\;;\;\;\;\;  \frac{\pa {\cal L}}{\pa Q^1} = 0 \; . \eeq Thus $Q^1$
is a cyclic coordinate and the dynamics can be reduced
\cite{arnold,marmo}. Clearly the change of coordinates defined by
(\ref{010}) is not unique. Usually a clever choice is very
important. It is possible that more than one vector field ${\bf X}$
is found. In this case, more than one symmetry exists.

\subsubsection{The case of Scalar-Tensor Gravity}
The above method can be used to seek for solutions in the dynamics
given by Lagrangian (\ref{actfr}). In particular, for flat FRW
metric, the field Lagrangian (\ref{actfr}) reduces to the
point-like Lagrangian
\begin{equation}
{\cal
L}=-3a\dot{a}^{2}F-3F_{,\Phi}\dot{\Phi}a^{2}\dot{a}+a^{3}\left(\frac{1}{2}\dot{\Phi}^2-U(\Phi)\right)-
Da^{-3(\gamma-1)}\,,
\label{e2}%
\end{equation}
where, for the sake of simplicity, we are considering only the
scalar field case (the generalization to the phantom field case is
obvious). The constant $D$ is related to the perfect-fluid matter
density, being $\rho_{\rm m}=D(a_0/a)^{3\gamma}$, where $1\leq
\gamma\leq 2$ defines the Zel'dovich range for the  equation of
state of standard matter. The above dynamical system
(\ref{stfe1}), (\ref{stfe2}), (\ref{klgord}) is immediately
deduced considering the energy condition  and the Euler-Lagrange
equations for (\ref{e2}). In the case of standard dust matter,
$\gamma=1$,  the last term in (\ref{e2}) reduces to an additive
constant.  Being $\{a,\Phi\}$ the configuration space of the
system, the problem is 2D and then the infinitesimal generator of
the Noether symmetry is
\begin{equation}
{\bf X}=\alpha \frac{\partial }{\partial {a}}+\beta \frac{\partial
 }{\partial {\Phi }}+\dot{\alpha}\frac{\partial }{\partial
 {\dot{a}}}+\dot{\beta}\frac{\partial }{\partial {\dot{\Phi}}},
\end{equation}
where $\alpha $ and $\beta $ are functions depending on $a$ and
$\Phi $, and
\begin{equation} \dot{\alpha}\equiv\frac{\partial
\alpha }{\partial
 a}\dot{a}+\frac{\partial
\alpha }{\partial \Phi }\dot{\Phi}\quad ;\quad
 \dot{\beta}\equiv\frac{\partial
\beta }{\partial a}\dot{a}+\frac{\partial \beta }{\partial \Phi
 }\dot{\Phi}.
\end{equation}
 The condition for the existence of a Noether symmetry is
$L_{\bf X}{\cal L}=0$. It, explicitly, gives an expression of second
degree in $\dot{a}$ and $\dot{\Phi}$, whose coefficients are zero
due to the fact they are considered to be linearly independent. Then
this set of coefficients gives rise to the following system of
partial differential equations \cite{cimento},
\begin{equation}
\alpha +2a\frac{\partial {\alpha }}{\partial
{a}}+a^{2}\frac{\partial
 {\beta}}{\partial {a}}\frac{F_{,\Phi }}{F}+a\beta \frac{F_{,\Phi }}{F}=0
\label{eq:i}
\end{equation}
\begin{equation}
\label{eq:iii} \left( 2\alpha +a\frac{\partial {\alpha }}{\partial
 {a}}+a\frac{\partial
\beta }{\partial \Phi }\right) F_{,\Phi }+aF_{,\Phi\Phi }\beta
 +2F\frac{\partial \alpha }{\partial \Phi }-\frac{a^{2}}{3}\frac{\partial{\beta}}{\partial {a}}=0
\end{equation}
\begin{equation}
3 \alpha -6 F_{,\Phi }\frac{\partial {\alpha }}{\partial
 {\Phi}}+2 a \frac{\partial {\beta }}{\partial
 {\Phi}}=0 \label{eq:ii}
\end{equation}
\begin{equation}
{U_{,\Phi} \over U} = - \frac{3 \alpha}{a \beta} \label{eq:iv0}
\end{equation}

Equation (\ref{eq:iv0}) can be re-written in the form

\begin{equation}
{U_{,\Phi} \over U} = m \cdot {F_{,\Phi } \over F} \label{eq:iv}
\end{equation}

where \begin{equation} m \equiv - \frac{3 \alpha}{a \beta} \frac{
F}{F_{,\Phi }} \label{defm}
\end{equation}

It is worth noting that Eq.(\ref{eq:iv})  is a relation between
the potential and the coupling and it exactly coincides with
(\ref{mdef}). Solving the above system means to find out the
explicit form of the set of functions $\{\alpha,\beta,F,U\}$. For
this purpose we consider the separation of variables,

\be \alpha = A_1 (a) A_2(\Phi) \label{defalpha} \ee

and

\be \beta = B_1 (a) B_2(\Phi) \label{defbeta} \ee Then from eq.
(\ref{eq:ii}) we get

\be \frac{B_1 a}{6 A_1}=-\frac{A_2}{4 B_2'}+\frac{A_2' F'}{2
B_2'}=C \label{separ1} \ee where $C$ is a separation constant. The
solution of (\ref{separ1}) is simple and we get

\be A_1=\frac{B_1 a}{6 C} \label{defA1} \ee

and

\be B_2'=-\frac{A_2-2 A_2'F'}{4C} \label{B2primedef} \ee Using
eqs. (\ref{defalpha}) and (\ref{defbeta}) on (\ref{eq:i}) we get

\be -\frac{a}{B_1}\frac{dB_1}{da}=\frac{3(A_2 F +2 C B_2
F')}{2(A_2 F +3 C B_2 F')}=-s \label{B1primedef} \ee where $ s $
is a separation constant. Hence, from eqs. (\ref{defA1}) and
(\ref{B1primedef}) we get

\be B_1=B a^s \label{B1} \ee and \be A_1=\frac{B}{6C}a^{s+1}
\label{A1} \ee where $B$ is a constant of integration. Also, from
(\ref{B1primedef}) we have

\be \frac{F'}{F}=-\frac{2s+3}{6C(s+1)}\frac{A_2}{B_2}
\label{FprimeoverF} \ee Using eqs. (\ref{B1}), (\ref{A1}) and
(\ref{eq:ii}) yields

\be 2 F A_2'+\left(A_2 (s+3)+6 C B_2'\right) F'-2 B_2 C \left(s-3
F''\right)=0 \label{eqFA2B2} \ee

Now we are left with three equations (\ref{B2primedef}),
(\ref{FprimeoverF}) and (\ref{eqFA2B2}) to solve for the three
unknown functions $A_2$, $B_2$ and $F$. Using eqs.
(\ref{B2primedef}) and (\ref{FprimeoverF}) for $B_2$ and $B_2'$ on
eq. (\ref{eqFA2B2}) we get

\be F A_2'+\frac{F'}{4} \left(A_2 (2 s+3)+6 A_2'
F'\right)+\frac{A_2 F (2 s+3) \left(s-3 F''\right)}{6 (s+1) F'}=0
\label{A2Fprime} \ee Also, using (\ref{FprimeoverF}) in
(\ref{B2primedef}) yields

\ba 2 A_2' F' \left(3 (s+1) \left(F'\right)^2+F (2 s+3)\right)+
\nn \\+ A_2 \left((s+3) \left(F'\right)^2-2 F (2 s+3) F''\right)=0
\label{A2Fprime1} \ea

Now, we can use equations (\ref{A2Fprime}) and (\ref{A2Fprime1})
to eliminate $A_2$ and $A_2'$ in favor of $F$,

\ba   F''&=& \frac{3 s (s+1) (s+2) F'^4}{(2 s+3) F^2} +
 \nn \\ &+& \frac{(s+1) \left(8 s^2+16 s+3\right) F'^2}{2 (2 s+3)
F}+\frac{s (2 s+3)}{3}  \label{FODE} \ea

Equation (\ref{FODE}) is nonlinear and its complete solution is an
elliptical integral of second kind which is not simple to handle.
However, an exact solution can be found to be of the form \be
F=\xi (\Phi -\Phi_0)^2 \label{defF1} \ee Using the ansatz
(\ref{defF1}) in (\ref{FODE}) we get that $\xi=-\frac{(2
s+3)^2}{24 (s+1) (s+2)}$ or $\xi=-\frac{1}{6}$, with the latter
corresponding to the conformal coupling \cite{Faraoni:1998qx}. 
Note also that the conformal coupling $\xi = -1/6$ corresponds to
a model where there is actually no scalar degree of freedom
despite the fact that there seems to exist one in the
parametrization (\ref{actfr}) (see \cite{Deser:1970hs}). The free
parameter $s$ has a physical meaning since it is connected to the
ratio of critical points ${\bar x}_1$ and ${\bar x}_3$ and to the
coupling. For this form of $F(\Phi)$ we can now determine $A_2$
from (\ref{A2Fprime}) or (\ref{A2Fprime1}) and $B_2$ from
(\ref{FprimeoverF}) and finally arrive at a solution for $\alpha$
and $\beta$. For the two values of $\xi$ and eqs. (\ref{A2Fprime})
and (\ref{A2Fprime1}) we get three degenerate solutions for
$\alpha$ and $\beta$, ie they correspond to the same form of the
potential $U$ (see eq.(\ref{eq:iv0})). The solutions are

\bea \alpha_1 &=& \frac{a^{s+1} A B (\Phi -\Phi_0)^{\frac{2 s
(s+2)}{2 s+3}}}{6 C} \\
\beta_1 &=& \frac{a^s A B (2 s+3) (\Phi -\Phi_0)^{\frac{2 s
(s+2)}{2 s+3}+1}}{12 C (s+1)} \label{aandb1} \eea

\bea \alpha_2 &=& \frac{a^{s+1} A B (\Phi -\Phi_0)^{\frac{s (2
s+3)}{2 (s+1)}}}{12 C (s+1)} \\ \beta_2 &=& \frac{a^s A B (2 s+3)
(\Phi -\Phi_0)^{\frac{s (2s+3)}{2 (s+1)}+1}}{12 C (s+1)}
\label{aandb2} \eea

\bea \alpha_3 &=& \frac{a^{s+1} A B (\Phi_0-\Phi )^s}{6 C} \\
\beta_3 &=& \frac{a^s A B (2 s+3) (\Phi_0-\Phi )^{s+1}}{12 C
(s+1)} \label{aandb3} \eea It is easy to show that for all three
cases we have \be m(s) = \frac{3 (s+1)}{2 s+3} \label{ms} \ee and
that from eq. (\ref{eq:iv}) we get \be U(\Phi)=U_0 (\Phi
-\Phi_0)^{\frac{6 (s+1)}{2 s+3}}=U_0  (\Phi
-\Phi_0)^{2m(s)}\label{potent} \ee where $U_{0}$ is a constant
determining the scale of the potential and is not directly
measurable, but can be rewritten in terms of observable parameters
like $H_0$, $q_0$, $\Omega_{\rm m }$. A noteworthy feature of eq.
(\ref{potent}) is that it exactly coincides with (\ref{VofPhi1}),
thus hinting towards a non-trivial physical content in this class
of scalar-tensor Lagrangians.

In order to find the solution to the field equations we need to
find the value of the constant of motion $\Sigma_0$ from eq.
(\ref{09}). This is, \be \Sigma_0 = \frac{b (2 s+3)^2} {(s+1)^2
(s+2) (s+3)} \frac{d(a^{s+3} (\Phi-\Phi_0)^{\frac{2 s (s+2)}{2
s+3}+2})}{dt} \label{sigma0}\ee

where $b=-\frac{A B}{48 C}$. Integration of (\ref{sigma0}) yields
\be \Phi-\Phi_0 = a^{-\frac{2 s+3}{2 s+2}} c^{1+\frac{3}{2 s}}
t^{\frac{2 s+3}{2 s^2+8 s+6}} \label{phiminusphi0} \ee where
$c=\left(\frac{(s+1)^2 \left(s^2+5 s+6\right) \Sigma_0 }{b (2
s+3)^2}\right)^{\frac{s}{(s+1) (s+3)}}$.

Plugging (\ref{potent}) and (\ref{phiminusphi0}) into
(\ref{stfe1}) we can get $a(t)$ and $\Phi(t)$:

\be a(t)=t^{\frac{s+2}{s+3}} \left(\frac{8 c s (s+1) (s+2) (s+3)^2
U_0 t^{\frac{s+6}{s+3}}}{(s+6) (2
s+3)^2}+a_0\right)^{1+\frac{1}{s}} \label{at}\ee

and

\ba (\Phi(t)-\Phi_0)^2 = c^{2+\frac{3}{s}} t^{-\frac{2
s+3}{s+3}}\cdot \nn ~~~~~~~~~~~~~~~~~~~~~~~
\\\left(\frac{8 c s (s+1) (s+2) (s+3)^2 U_0
t^{\frac{s+6}{s+3}}}{(s+6) (2 s+3)^2}+a_0\right)^{-2-\frac{3}{s}}
\label{phioft}\ea where $a_0 $ is an integration constant.

Due to the structure of the above general solution, the cases
$s=0,-1,-3/2,-2$ and $s=-3$ have to be considered apart. The
solutions for $s=0$ and $s=-3$ correspond to the minimal coupling
where $F=F_0$ and $U=\Lambda$ and to the quartic potential case
where $F\sim\Phi^2$ and  $U\sim \Phi^4$. In these situations, the
solutions assume oscillating or exponential behavior (for a
discussion  see \cite{Marek1995,cimento,Dem}).

\subsection{Numerical results}
As final step, in order to confirm the validity of our analysis, we
perform a numerical evolution of the dynamical system to compare the
form of the potential $U$ and coupling $F$ of eqs.(\ref{FofPhi}) and
(\ref{VofPhi1}) (found also by the Noether Symmetry Approach), with
the corresponding form obtained from the numerical analysis. The
steps involved in the comparison are the following:

\begin{itemize}
    \item Numerically solve the dynamical system (\ref{au1}), (\ref{au4}), (\ref{au2}),
    (\ref{au3}) and obtain $x_1(N)$, $x_2(N)$, $x_3^2(N)$ and
    $x_4(N)$. Integrate equation (\ref{x1}) to get
    \be F(N)=F_0 e^{-\int_{N_{min}}^N x_1(N')dN'} \label{fnnum} \ee
    and use eq. (\ref{x3}) in the form \be \Phi'(N)^2=6 x_3^2(N) F(N) \label{PhipN} \ee to obtain $\Phi(N)$. The resulting
    form of $F(\Phi)$ in both the numerical (red continuous line) Eq.(\ref{fnnum}) and its analytical
    approximation (blue dotted line) Eq.(\ref{FofPhi}) is shown in Fig. 4.

    \item Use equation (\ref{x2}) to obtain $U(N)$ numerically (red continuous line in Fig. 5)
    \be U(N)=3 x_2(N) F(N) H(N)^2 \label{Vnum} \ee and compare
    with the analytical form (blue dotted line in Fig. 5) of Eq.(\ref{VofPhi1}).
\end{itemize}

\begin{figure*}[pt!]
\begin{center}
\rotatebox{0}{\resizebox{.85\textwidth}{!}{\includegraphics{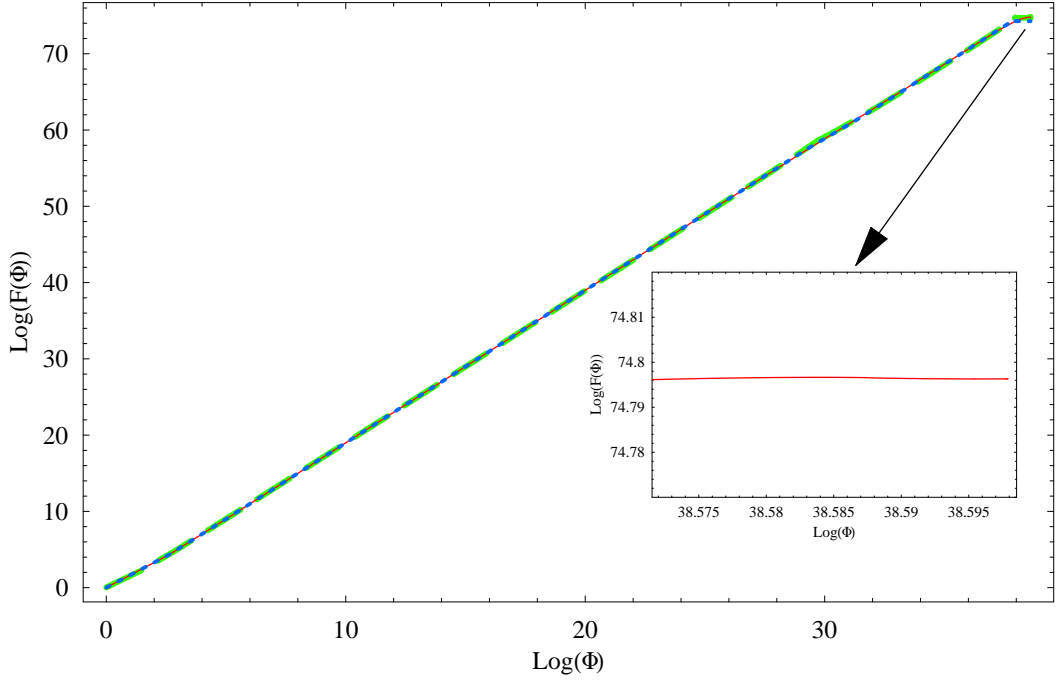}}}
\vspace{0pt}{ \caption{The form of $log(F(\Phi))$ in the numerical
reconstruction (red continuous line), its analytical approximation
(blue dotted line) and a fit of the numerical reconstruction using
Eq.(\ref{fnnum}) (green long-dashed line). The agreement between
the three approaches is very good. The reason for the existence of
the small plateau, see the zoomed region, is that as the system
evolves towards the deSitter era the potential $F(\Phi)$
``freezes" much faster than the field $\Phi$. }} \label{fig4}
\end{center}
\end{figure*}

\begin{figure*}[pb!]
\rotatebox{0}{\resizebox{.8\textwidth}{!}{\includegraphics{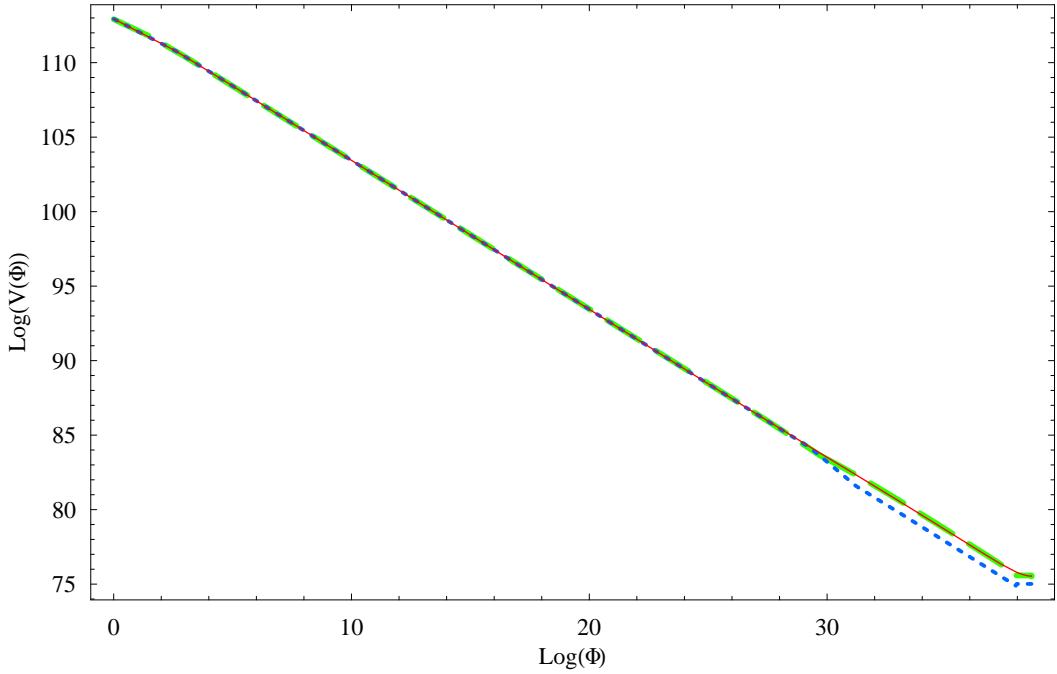}}}
\vspace{0pt}{ \caption{The form of $log(V(\Phi))$ in the numerical
reconstruction (red continuous line), its analytical approximation
(blue dotted line) and a fit of the numerical reconstruction
(green long-dashed line). The agreement between the three
approaches is very good. The potential exhibits a small plateau in
the deSitter era for the same reason as $F(\Phi)$ (see caption of
Fig. 4)}} \label{fig5}
\end{figure*}

In the methodology mentioned above we used the exact coefficients
$\bar{x}_1$, $\bar{x}_2$, $\bar{x}_3^2$ and $\bar{x}_4$ from Table
I in each era for the analytic forms (\ref{FofPhi}) and
(\ref{VofPhi1}). Also, the initial conditions used in the
numerical evolution were $F(N=-30)=\Phi(N=-30)=1$. As another
test, we fitted the numerically obtained $F(N)$ and $U(N)$ from
eqs.(\ref{fnnum}) and (\ref{Vnum}) respectively to obtain the
coefficients of the analytic forms (\ref{FofPhi}) and
(\ref{VofPhi1}) and found that the results were in good agreement
(see Figs. 4 and 5 the green long-dashed lines). A noteworthy
feature of Figs. 4 and 5 is a small plateau that appears during
the deSitter era (see the zoomed region in Fig. 4). The reason for
the existence of the plateau is that as the system evolves towards
the deSitter era the coupling $F(\Phi)$ ``freezes", since
$x_1(N_{dS})\rightarrow 0$ (see eq. (\ref{x1}) and Fig. 2a), much
faster than the field $\Phi$.

\section{Conclusions and Outlook}
We have reconstructed the form of the gravitational coupling
$F(\Phi)$ and the potential $U(\Phi)$ of scalar-tensor quintessence
by demanding that it reproduces a $\Lambda CDM$ cosmic history
through the radiation ($w_{eff}=\frac{1}{3}$), matter ($w_{eff}=0$),
and deSitter ($w_{eff}=-1$) eras. We have found that apart from the
usual general relativistic solution with a constant coupling
$F(\Phi)=F_0$ and potential $U(\Phi)=U_0$ (corresponding to Newton
and cosmological constants), there is another consistent solution
which reproduces the same cosmic history. According to this solution
\ba F(\Phi)&=&\xi (\Phi - C)^2 \label{fphi1}
\\ U(\Phi)&\sim& F(\Phi)^m \label{uphi1}\ea where $m$, $C$ are arbitrary
constants ($m$ however is negative). In this new solution the 
`radiation', `matter' and `deSitter' expansion rates, for the
`attractor' trajectory shown in Figs 2 and 3, is dominated by dark
gravity through all epochs. This is indeed a potential problem for
this type of trajectories but it could also be a potential
blessing since this type of solutions have the correct expansion
rate at all epochs without the use of dark matter or dark energy.
A proper test of these models for detailed comparison with
observations would require analysis of large scale structure
formation (analysis of evolution of \textit{perturbations}). Such
an analysis is beyond the aims of our present analysis but it is
an interesting extension of this project.

We should also stress that not only the attractors but all the
fixed points we found are potentially interesting because as
discussed above a more physical stability analysis would fix
$F(N)$ and allow $H(N)$ to vary thus introducing and eliminating
instability modes. Thus the physically interesting part of our
analysis is the actual values of the fixed points and not their
stability which could change if $H(N)$ were allowed to vary.
On the other hand, going back to the considerations developed in
Sect.II regarding the number of parameters beside $m$, we have to
stress again that if we had used the parameter $n\sim F_{,\Phi}/F$
in our analysis (thus fixing $F(\Phi)$ and allowing $H(N)$ to
vary), the autonomous system would be very different and,
depending on the values of $n$, the stability and the form of
$H(N)$ would correspondingly vary. An alternative approach could
be to fix $F$ and $U$, as in Eqs.(\ref{fphi1}) and (\ref{uphi1}),
and allow $H(N)$ to vary about a the $\Lambda$CDM background,
along the lines of Ref. \cite{SanteLeach}. Very likely, this
approach could provide a comprehensive stability analysis of the
$\Lambda$CDM model  since it  involves also the proper fluctuation
modes of $H(N)$. This extension is out of the lines of this work
and will be faced in a forthcoming paper.

Another point is that phantom behavior can be easily realized in
scalar tensor theories, see Refs \cite{Gannouji,Martin:2005bp} for
a discussion, and this is an attractive feature of these theories.
In fact we could have chosen to reconstruct different forms of
$H(N)$ giving late time phantom behavior thus deriving different
forms of $F(N)$ at late times with different fixed points. This
late time reconstruction has been undertaken in Refs.
\cite{Nesseris:2006er} and \cite{Tsujikawa:2005ju}. However,
clearly the early times behavior of the reconstructed $F(N)$ would
be unchanged even in the phantom case. Fixing $F(\Phi)$ could also
lead to phantom behavior but we would have to guess a proper form
of $F$.

A completely independent way that can also lead to the form of the
new solution presented here is obtained by imposing maximal Noether
symmetry on the scalar-tensor Lagrangian. We have demonstrated that
imposing such a symmetry leads uniquely to exactly the same form of
potentials as (\ref{fphi1}) and (\ref{uphi1}). It also leads to a
conserved charge $\Sigma_0$ which allows the derivation of exact
solutions for the evolution of the scale factor $a(t)$ and the
scalar field $\Phi(t)$.

This intriguing coincidence of the two approaches hints towards a
non-trivial physical content in this class of scalar-tensor
Lagrangians. It is therefore important to study the evolution of
cosmological perturbations in this class of models in order to test
them using Large Scale Structure and CMB observations.

\vspace{8 mm}

\noindent {\bf Acknowledgements:} The authors would like to thank
Gilles Esposito-Farese and David Polarski for helpful suggestions
and discussions. This work was supported by the European Research
and Training Network MRTPN-CT-2006 035863-1 (UniverseNet). S.N.
acknowledges support from the Greek State Scholarships Foundation
(I.K.Y.).

\end{document}